\documentstyle[12pt,epsfig,amssymb]{article}
\def\ltsim{\lower3pt\hbox{$\, \buildrel < \over \sim \, $}}
\def\gtsim{\lower3pt\hbox{$\, \buildrel > \over \sim \, $}}
\def\be{\begin{equation}}
\def\ee{\end{equation}}
\def\ba{\begin{eqnarray}}
\def\ea{\end{eqnarray}}
\def\ga{\mathrel{\raise.3ex\hbox{$>$\kern-.75em\lower1ex\hbox{$\sim$}}}}
\def\la{\mathrel{\raise.3ex\hbox{$<$\kern-.75em\lower1ex\hbox{$\sim$}}}}

\begin{document}
\baselineskip=16pt
\begin{titlepage}
\rightline{UG--FT--140/02}
\rightline{CAFPE--10/02}
\rightline{hep-ph/0207319}
\rightline{July 2002}  
\begin{center}

\vspace{1.5cm}

\large {\bf 
Light neutrino propagation in matter \\[2mm]
without heavy neutrino decoupling 
\footnote{Dedicated to Stefan Pokorski on his 60th birthday}}
\vspace*{10mm}
\normalsize

{\bf F. del Aguila $^a$ \footnote{faguila@ugr.es} and 
M. Zra\l{}ek $^b$ \footnote{zralek@us.edu.pl}}

\smallskip 
\medskip 
{\it $^a$ Departamento de F\'\i sica Te\'orica y del Cosmos and 
Centro Andaluz de F\'\i sica de Part\'\i culas Elementales, 
Universidad de Granada, E-18071 Granada, Spain} \\[2mm]
{\it $^b$ Institute of Physics, University of Silesia,
Uniwersytecka 4, 40-007~Katowice, Poland}

\smallskip  

\vskip0.8in 
\end{center}
 
\centerline{\large\bf Abstract}
We review the propagation of light neutrinos in matter assuming 
that their mixing with heavy neutrinos is close to present experimental 
limits. The phenomenological implications of the non-unitarity of the light 
neutrino mixing matrix for neutrino oscillations are discussed. 
In particular we show that the resonance effect in neutrino propagation 
in matter persists, but for slightly modified values of the parameters 
and with the maximum reduced by a small amount proportional to the mixing 
between light and heavy neutrinos squared. 

\noindent
\vfill
\begin{center}
 \\[0.2mm]
\end{center}
\end{titlepage}

\section{Introduction}
There is convincing evidence for neutrino masses and mixing, 
being at least three light neutrinos  
with masses $\lesssim 2.2$ eV \cite{PDB,Reviews}. In fact LEP has measured
the number of light standard neutrinos $\rm N_\nu = 2.994\pm 0.012$, 
excluding new ones with masses below $\sim \frac{M_Z}{2}$.
Light neutrinos with small couplings, sterile, and heavy ones are 
not ruled out, although there are astrophysical and cosmological 
constraints on their masses, nature and decay lifetimes \cite{Mohapatra}.

In order to describe their interactions it is usually assumed 
that the mixing of the three light standard neutrinos is given by a unitary 
matrix, and then that their mixing with heavy neutrinos  
(and the lack of unitarity) is negligible. In practice this is the case for 
see-saw models \cite{GellMann79}.
Indeed, if for the sake of discussion we assume only one light neutrino, 
the mass matrix 
\begin{equation}
\left( \matrix{ 0 & v \cr
                        v & M} \right)
\end{equation}
requires a very heavy Majorana mass, typically 
of the order of the unification scale $M \sim 10^{15}$ GeV, 
to generate light masses $m \sim \frac{v^2}{M}\sim $ eV, 
with $v \sim 250$ GeV the electroweak vacuum espectation value. 
As a consequence the mixing between light and heavy neutrinos 
$\frac{v}{M} \sim 10^{-25}$, and thus completely negligible.
The numerical problem can be improved introducing a small Yukawa 
coupling $\lambda $ ($v\rightarrow \lambda v$ everywhere), but not evaded.
However, one can write down models where the light masses and  
mixings are not correlated, allowing in principle for observable 
non-decoupling effects proportional to the mixing between 
light and heavy neutrinos. 
In particular in the general $2\times 2$ case 
\begin{equation}
\left( \matrix{ a & \lambda v \cr
                 \lambda v & M} \right)
\end{equation}
the light mass $m\sim a -\frac{\lambda ^2v^2}{M}$ can vanish and 
at the same time the mixing $\sim \frac{\lambda v}{M}$ can be relatively 
large if we fine tune $a$. 
This scenario can be made more natural adding new degrees of freedom.
For example, one can write models with two heavy neutrinos 
$N$ and $N'$ per family and an effective approximate symmetry 
$L_{\nu} + L_N - L_{N'}$ implying a mass matrix of the form
\begin{equation}
\left( \matrix{ 0 & 0 & \lambda v \cr
                0 & 0 & M \cr
                \lambda v & M & 0} \right),
\label{MNN}
\end{equation}
where a large singlet vacuum expectation value $M$ gives a 
Dirac mass to the heavy neutrinos, whereas the light neutrino is 
massless and the mixing between the light and heavy neutrinos 
$\frac{\lambda v}{M}$ arbitrary. 
This is similar to the light neutrino mass matrix texture obtained 
imposing the lepton number symmetry $L_{\nu_e} - L_{\nu_\mu} - L_{\nu_\tau}$
\cite{Barbieri98}. 
Eq. (\ref{MNN}) generalizes to three families trivially 
but leaves three massless neutrinos.
If we want to give them a small mass, 
we can introduce a Majorana mass $m'\ll M$ 
for the heavy neutrino $N$, violating 
the approximate symmetry and inducing a 
light neutrino mass $m \sim m'\frac{\lambda ^2v^2}{M^2}$. 
An alternative 
way is to assume that there exists a much heavier Majorana fermion 
which through the see-saw mechanism gives a very small mass to the 
light neutrino, violating 
also the approximate symmetry (up-left entry), and mixes very little. 
At any rate, it seems necessary 
in order to have small enough neutrino masses and at the same time 
a relatively large mixing between light and heavy neutrinos, 
that both have different origin. Models with extra dimensions 
can do the job \cite{Arkani02,Dienes99}. 
A neutral fermion living in the bulk can reduce to 
a massless right-handed neutrino plus a tower of heavy Kaluza-Klein 
modes. Then as pointed out in Ref. \cite{Dienes99} 
after the electroweak symmetry 
breaking the new fermions can mix with a standard neutrino and 
give a massless mode with a relatively large mixing $\sim \lambda v R$,
where $R$ is the compactification radius. In this case the truncation 
of the Kaluza-Klein tower can also generate a tiny neutrino 
mass $\sim \frac{\lambda ^2v^2}{M_s}$, with $M_s$ 
the mass scale of the underlying (string) theory where the infinite 
Kaluza-Klein tower is truncated. 
  
If one assumes a relatively large departure of the unitary mixing 
among light neutrinos, one must wonder about  
possibly large contributions 
to rare leptonic processes, {\it e.g.} $\mu \rightarrow e \gamma, 
\mu \rightarrow e e\bar e, Z \rightarrow e \bar \mu, ...$.
As no such decays have been observed, relatively stringent 
bounds on the mixing between light and heavy neutrinos 
and the heavy masses can be derived \cite{Aguila83}. 
In the following independently of their origin we discuss the effects 
of non-decoupled heavy neutrinos in light neutrino physics, in particular 
in neutrino oscillations. 

New contributions to processes involving only the known fermions as 
initial and final states are typically proportional to the square of the 
mixing between light and heavy neutrinos, and then small and difficult 
to observe. This makes processes forbidden in the absence of such a 
mixing particularly interesting. Prime examples are the lepton number 
violating processes involving charged leptons and CP violating 
neutrino oscillations. If the angle mixing 
the electron and tau neutrinos is not small but negligible, no CP 
violating neutrino oscillation is observable if the light neutrino 
mixing matrix is unitary. This does not need to be the case if 
the light neutrinos mix with heavy ones making the light neutrino 
mixing matrix non-unitary. Hereafter we will discuss this 
possibility following closely Ref. \cite{Bekman02} but sticking mainly to 
the eigenmass basis description of neutrino oscillations. 
In Section 2 we introduce the neutrino 
bases convenient for describing neutrino propagation in matter 
\cite{Wolfenstein78}, 
which we review in Section 3. In Section 4 we study 
the case of two neutrinos propagating 
in unpolarised, isotropic and neutral matter, and in Section 5 
we calculate the corrections to the resonance effect in 
neutrino oscillations. 
Section 6 is devoted to conclusions.
   
The mixing with heavy neutrinos implies the loss of unitarity 
of the Maki-Nakagawa-Sakata (MNS) mixing matrix \cite{Maki62} 
describing the charged current interactions. The same 
happens if the observed charged leptons mix with new heavy 
ones \cite{Branco02}. The phenomenological consequences are also 
similar.  
Both cases are explicit examples of the Standard Model (SM) 
extensions parametrised in Ref. \cite{Gonzalez01}.
Present limits on rare processes postpone any observation 
of these effects in neutrino oscillations to 
$\nu$ factory experiments \cite{NUfactory}.
 
\section{Neutrino eigenstates}
Let us assume that there are three light active, $n_s$ 
light sterile and $n_R$ heavy neutrinos. So the mass matrix 
has dimension $n = 3 + n_s + n_R$, being diagonalised by a 
unitary matrix 
\begin{equation} 
U^T_\nu M U_\nu = (M_\nu)_{\mathrm {diag}}\equiv  diag (m_1 m_2 ... m_{3+n_s} 
M_1 ... M_{n_R}),
\end{equation}
where
\begin{equation}
 U_\nu = \left( \matrix{ {\cal U} & {\cal V} \cr
                        {\cal V}' & {\cal U}'} \right),
\label{Unu}
\end{equation}
with the $(3+n_s)\times(3+n_s)$ matrix 
${\cal U}$ ($n_R\times n_R$ matrix ${\cal U}'$) 
describing the mixing among the light (heavy) neutrinos and 
the matrices ${\cal V}$ and ${\cal V}'$ parametrising the 
mixing between the light and heavy neutrinos. 
Thus, the flavour eigenstates are linear combinations of the 
mass eigenstates 
\begin{equation} 
| \nu_\alpha > =  \sum\limits_{i=1}^n
(U_{\nu}^{\ast})_{\alpha i} | \nu_i >  =
\sum\limits_{i=1}^{3+n_{s}} {\cal U}_{\alpha i}^\ast | \nu_i > +
\sum\limits_{i=3+n_{s}+1}^{3+n_{s}+n_{R}} {\cal  V}_{\alpha
i}^\ast | \nu_i >, \label{ch} 
\end{equation}
with $\alpha = 1, 2, 3$ standing for $e, \mu , \tau $, 
respectively. 
In the charged lepton mass eigenstate basis the first three rows of 
$U_\nu$ parametrise the charged and neutral current interactions, the 
corresponding Lagrangians being
\begin{equation}
L_{CC}=\frac{e}{2 \sqrt{2} \sin \theta_W}
\sum\limits_{\alpha=e,\mu,\tau} \sum\limits_{i=1}^n
\bar{l}_{\alpha} \gamma^{\mu} \left( 1 - \gamma_5 \right)
(U_{\nu})_{\alpha i} \nu_i W_\mu^- + h.c. \label{cc} 
\end{equation}
and
\begin{equation}
\begin{array}{rl}
L_{NC}\ = & \frac{e}{4  \sin \theta_W \cos \theta_W}
\left\{ \sum\limits_{i,j=1}^n \bar{\nu}_{i} \gamma^{\mu} \left( 1 -
\gamma_5 \right) \Omega_{ij} \nu_j \ + \right. \\
& \left. 2 \sum\limits_{f=e,p,n}   \bar{f} \gamma^{\mu} \left[
T_{3f} \left( 1 - \gamma_5 \right) -2 Q_f \sin^2 \theta_W
\right]  f \right\} Z_\mu \label{nc}  , 
\end{array}
\end{equation} 
where $T_{3f}$ and $Q_f$ are the third component of the weak isospin 
and the charge of the fermion $f$, respectively, and 
$\Omega_{ij}=\sum_{\alpha=e,\mu,\tau} 
{(U_\nu)}_{\alpha i}^\ast {(U_\nu)}_{\alpha j}$.
The non-observation of SM deviations (except for 
neutrino oscillations) bounds the new interactions. Universality 
sets limits on the diagonal elements of 
\begin{equation}
\omega_{\alpha \beta} \equiv (\cal V \cal V ^\dagger)_{\alpha \beta} 
= \delta_{ \alpha \beta} -
\left( {\cal U}{\cal U}^\dagger \right)_{ \alpha \beta}
\ ,
\end{equation}
and the off-diagonal ones are mainly constrained by 
the non-observation of the 
lepton number violating processes $\mu \rightarrow e \gamma, 
\mu \rightarrow e e\bar e, Z \rightarrow e \bar \mu, ...$ 
\cite{Aguila83} 
\begin{equation}
\begin{array}{c}
\omega_{ee} < 0.0054, \;\;\; 
\omega_{\mu \mu} < 0.0096, \;\;\; 
\omega_{\tau \tau} < 0.016, \\ 
| \omega_{e \mu} | < 0.0001, \;\;\; 
| \omega_{\mu \tau} | < 0.01
\label{eps1}
\end{array}
\end{equation}
(assuming no model dependent cancellation).

Future experiments will improve these bounds or detect new 
effects. In neutrino oscillations with low energy production and 
detection processes and heavy neutrinos not propagating large 
distances the effective flavour states are obtained truncating 
Eq. (\ref{ch})   
\begin{equation}
| \tilde{\nu}_\alpha> = \lambda_{\alpha}^{-1}
\sum\limits_{i=1}^{3+n_s} {\cal U}^\ast_{\alpha i} | \nu_i > \equiv
\sum\limits_{i=1}^{3+n_s} \tilde{{\cal U}}_{\alpha i}^\ast | \nu_i >,
\label{real} 
\end{equation}
where we have also conventionally included the normalization factor 
$\lambda_{\alpha}=\sum_{i=1}^{3+n_s} |{\cal U}_{\alpha
i}|^2=\sqrt{1-\omega_{\alpha \alpha}}$.
These states do not need to be orthogonal 
\begin{equation} 
< \tilde{\nu}_\alpha | \tilde{\nu}_\beta > =
\left( \lambda_{\alpha}\lambda_{\beta} \right) ^{-1}
\left( \delta_{\alpha \beta} - \omega_{\alpha \beta}\right) ,
\label{nort} 
\end{equation}
reading in the flavour basis 
\begin{equation}
| \tilde{\nu}_\alpha> = \lambda_{\alpha}^{-1}
\left(| \nu_\alpha> - \sum\limits_{\beta=1}^{3+n_s} 
\omega_{\beta\alpha} | \nu_\beta > +  
\sum\limits_{\beta=3+n_s+1}^{n} \left({\cal V}'{\cal U}^\dagger \right) 
_{\beta \alpha} | \nu_\beta > \right) . 
\end{equation}
As an example, let us consider neutrino production in the 
charged current process $l^{-}_{\alpha} X \to \nu_{\beta} Y$.
If the available mass 
\begin{equation}
\Delta m_{\beta} = \Delta \left( \sqrt 
{(E_{l_{\alpha}^{-}} + E_X - E_Y )^2 - 
(\vec p_{l_{\alpha}^{-}} + \vec p_X - \vec p_Y )^2} \right)
\end{equation}
is much smaller than the heavy neutrino masses but much larger 
than the light ones, these will be produced coherently and 
the amplitude
\begin{equation}
\begin{array}{c}
A(l^{-}_{\alpha} X \to \tilde{\nu}_{\beta} Y) =
\lambda_{\beta}^{-1} \sum\limits_{i=1}^{3+n_s} {\cal U}_{\beta i} 
A(l^{-}_{\alpha} X \to \nu_i Y)  \\
\simeq \lambda_{\beta}^{-1} \sum\limits_{i=1}^{3+n_s} {\cal U}_{\beta i}
{\cal U}^{*}_{\alpha i}  A^{SM}(l^{-}_{\alpha} X \to \nu_{\alpha} Y) \\
= \lambda_{\beta}^{-1} (\delta_{\beta \alpha} - \omega_{\beta \alpha})
 A^{SM}(l^{-}_{\alpha} X \to \nu_{\alpha} Y) ,
\label{lambda1}
\end{array}
\end{equation}
where $A^{SM}(l^{-}_{\alpha} X \to \nu_{\alpha} Y)$ is the SM 
amplitude for massless neutrinos.
In particular 
\begin{equation}
\sigma (l^{-}_{\alpha} X \to \tilde{\nu}_{\alpha} Y)
\simeq \lambda_{\alpha}^2 
\sigma ^{SM}(l^{-}_{\alpha} X \to \nu_{\alpha} Y) .
\label{lambda2}
\end{equation}

\section{Neutrino propagation in matter}
Similarly to the case of photons the coherent scattering of light 
neutrinos in a medium modifies their properties. In the first case 
it gives the index of refraction of light, and for neutrinos 
it modifies their effective masses changing substantially 
their oscillation pattern, also showing resonance 
phenomena eventually \cite{Wolfenstein78}.  
The coherent neutrino scattering is described by a four-fermion 
Hamiltonian 
\begin{equation} 
H_{\mathrm {int}}^f(x)=\frac{G_F}{\sqrt 2} \sum\limits_{i,k=1}^{3+n_{s}}
\sum\limits_{a=V,A} \left[ \bar{\nu}_k \Gamma_a \nu_i \right] \left[
\bar{f} \Gamma^a \left( g_{fa}^{ki}+{\bar{g}}_{fa}^{ki}
\gamma_5 \right) f \right] ,
\label{Hint} 
\end{equation} 
where
$\Gamma_{V(A)}=\gamma_\mu (\gamma_\mu \gamma_5)$ and $f$ stands for the 
type of matter, electrons $e$ and nucleons $p,n$. 
This Hamiltonian and the couplings
$g_{fa}^{ki}$ and ${\bar{g}}_{fa}^{ki}$ can be calculated from
Eqs. (\ref{cc},\ref{nc}) \cite{Bekman02}. 
The Feynman diagrams are drawn in Fig. 1. 
\begin{figure}[ht]
\epsfig{figure=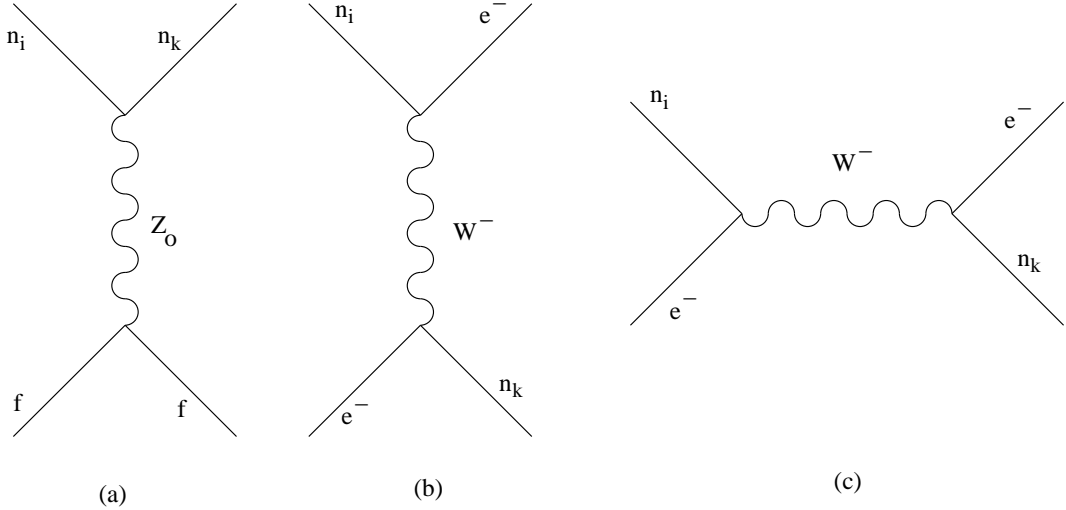, width=15 cm}
\caption{Feynman diagrams for neutrino 
scattering in matter. All three diagrams
contribute to neutrino-electron  scattering $n_i+e^- \to n_k+e^-$
$(f=e)$, but only diagram (a) contributes
to  neutrino-nucleon  scattering $n_i+f \to n_k+f$ $(f=p,n)$.} \label{diag}
\end{figure}
\begin{equation}
\begin{array}{cl}
g_{eV}^{ki} =-{\bar{g}}_{eA}^{ki}=&{\cal U}_{ek}^\ast {\cal
U}_{ei} +
\rho \Omega_{ki} \left( -\frac{1}{2}+ 2 \sin^{2} \theta_W \right), \nonumber \\
{\bar{g}}_{eV}^{ki}=-g_{eA}^{ki}=&-{\cal U}_{ek}^\ast {\cal
U}_{ei}+\frac 12 \rho \Omega_{ki}, \nonumber \\
g_{fV}^{ki} =-{\bar{g}}_{fA}^{ki}=&\rho \Omega_{ki} 
\left( T_{3f} - 2 Q_f \sin^{2}\theta_W \right), \nonumber \\
{\bar{g}}_{fV}^{ki}= - g_{eA}^{ki}=&-\rho \Omega_{ki} T_{3f},
\end{array}
\label{const}
\end{equation}
where $f=p,n$ and
\begin{equation}
\rho=\frac{M_W^2}{M_Z^2 \cos^2 \theta_W},\ T_{3p}=-T_{3n}=1/2,\ Q_p=1,\ Q_n=0.
\end{equation}
Then, the interaction Hamiltonian for a $\nu_i$ of momentum $\vec k$ and 
helicity $\lambda$ propagating in matter produces a $\nu_k$ with the same 
momentum and helicity is 
\begin{equation}
\begin{array}{rl}
H_{ki}^{\mathrm {int}}  = \ \sum\limits_{f}
& \int _{V=1} d^3x \frac {1}{N_f}
\sum\limits_{\vec s} \int  
\frac{d^3p}{\left( 2 \pi \right)^3} \rho_f ( \vec p, 
\vec s ) \\
& < \nu_k\ \vec k \ \lambda |  < f \ \vec p\  \vec s | 
H_{\mathrm {int}}^f (x) | f\ \vec p\  \vec s> 
| \nu_i\ \vec k \ \lambda > ,
\end{array}
\label{Hintt}
\end{equation}
where $\rho_f ( \vec p , \vec s )$ is the distribution function for the 
background fermions of type $f$, momentum $\vec p $ and spin $\vec s$ 
normalized to give the number of fermions per unit volume
\begin{equation} 
N_f \equiv \sum\limits_{\vec s} \int \frac{d^3p}{\left( 2 \pi
\right)^3} \rho_f ( \vec p , \vec s ) .  
\label{norm} 
\end{equation}
Assuming that the neutrinos are relativistic  
$\vec k ^2 \simeq E_{i,k} ^2 \gg m^2_{i,k}$ and 
using Eqs. (\ref{Hint},\ref{const}) we can write
\begin{equation}
\begin{array}{cl}
H_{ki}^{\mathrm {int}} (\vec k \ \lambda = -1) = &
- \left[ H_{ki}^{\mathrm {int}} (\vec k \ \lambda = +1) \right] ^\ast = \\
& \sqrt 2 G_F \sum\limits_{f} N_f 
\left[
g_{fV}^{ki} \left( 1 - \langle 
\frac{\vec{k} \cdot \vec{p}}{| \vec{k} | E_f} \rangle \right)\right. \\
&\left. + \ {\bar{g}}_{fV}^{ki} \left(
\langle \frac{\vec{p} \cdot \vec{s}}
{E_f} \rangle - m_f \langle \frac{\vec{k} \cdot \vec{s}}
{| \vec{k} | E_f} \rangle - 
\langle \frac{(\vec{k} \cdot \vec{p})(\vec{p} \cdot \vec{s})}
{| \vec{k} | (m_f + E_f)} \rangle
\right) 
\right] ,
\end{array}
\end{equation}
where $E_f = \sqrt{m_f^2 + \vec p^2}$ and $m_f$ are the energy and 
mass of the $f$ fermion, respectively, 
\begin{equation}
<z>\equiv\frac{1}{N_f} \sum\limits_{\vec s} \int  
\frac{d^3p}{\left( 2 \pi \right)^3} \rho_f ( \vec p, 
\vec s )  z ( \vec p, \vec s ) ,
\end{equation}
and $\lambda = - 1 \ ( + 1 )$ stands for the helicity of the 
Dirac (antineutrinos) and Majorana neutrinos.
This Hamiltonian enters the evolution equation for light neutrinos 
(expanding to first order in $H^{\mathrm {int}}$)
\begin{equation}
{\mathrm i} 
\frac{d}{dt} \psi_k(\vec k \ \lambda , t) = 
\sum\limits_{i=1}^{3+n_s} H_{ki}^{\mathrm {eff}} \psi_i(\vec k \ \lambda, t) ,
\label{s1}
\end{equation}
with $\psi_k(\vec k \ \lambda , t ) = 
< \nu _k \ \vec k \ \lambda | \psi (t) > $ and 
\begin{equation}
H_{ki}^{\mathrm {eff}}=
\frac{\Delta m_{i1}^2}{2|\vec k|} \delta_{ki}
+H_{ki}^{\mathrm {int}} (\vec k \ \lambda )\; .
\label{h1} 
\end{equation} 
As usual, we have removed the diagonal pieces of the effective 
Hamiltonian for they give global unobservable phases in 
neutrino oscillations. In particular 
$\Delta m^2_{i1}\equiv m_i^2-m_1^2$.
With these equations one can evaluate the different probability 
amplitudes. We apply them to a simple 
example in next Section.

\section{Propagation in an unpolarised, isotropic
 and electrically neutral medium}
Let us assume that for each fermion type $f$ matter is unpolarised 
$\langle \vec{s}\rangle = 0$ and isotropic 
$\langle \vec{p}\rangle = 0$, and as a whole  
electrically neutral $N_e = N_p \neq N_n$. In this case 
the interaction Hamiltonian is 
momentum and helicity independent
\begin{equation}
\begin{array}{rl}
H_{ki}^{\mathrm {int}} (\vec k \ \lambda) & = 
\sqrt 2 G_F \left[ N_e (g_{eV}^{ki}+g_{pV}^{ki}) 
 + N_n g_{nV}^{ki} \right] \\
& = \sqrt 2 G_F \left( N_e {\cal U}^*_{ek}{\cal U}_{ei}
 - \frac{1}{2} \rho  \Omega _{ki} N_n \right) .
\end{array}
\end{equation}
For constant density the evolution equation (\ref{s1}) can be easily 
solved diagonalizing the effective (hermitian) Hamiltonian 
\begin{equation}
\begin{array}{rl}
 H^{\mathrm {eff}}_{ki} &=  \frac{1}{2|\vec k|} \left(
\begin{array}{ccccc}
0 & 0 & 0 & 0 & \cdots \\ 0  & \Delta m_{21}^2 & 0 & 0 
& \cdots \\ 0 & 0 & \Delta m_{31}^2 & 0 & \cdots \\
0& 0& 0& \Delta m_{41}^2 & \cdots \\
 \vdots & \vdots & \vdots & \vdots & \ddots 
\end{array}
\right) _{ki} \\
&\\
&+ \sqrt{2} G_F \sum\limits_{\alpha ,\beta =1}^{3}
{\cal U}^\dagger _{k \alpha}  \left(
\begin{array}{ccc}
 (N_e-\frac{N_n}{2})  & 0 & 0  \\ 
0  & -\frac{N_n}{2} & 0 \\ 
0  & 0 & -\frac{N_n}{2} 
\end{array}
 \right) _{\alpha \beta} {\cal U} _{\beta i} \\
&\\
&= \frac{1}{2|\vec k|}\sum\limits_{j=1}^{3+n_s}
 W_{kj}^\dagger \tilde m_j^2 W_{ji},
\end{array}
\end{equation}
where $\tilde m_j^2$ are the effective (real) masses and $W_{ji}$ 
the diagonalising (unitary) matrix giving the effective mass neutrinos 
as linear combination of the vacuum mass ones.
Hence 
\begin{equation}
\begin{array}{rl}
A_{\tilde \nu _\alpha \to \tilde \nu _\beta} (L)&\equiv
< \tilde{\nu}_\beta (0) | \tilde{\nu}_\alpha (t=L) > \\
&=\lambda _\beta ^{-1} \lambda _\alpha ^{-1} 
\sum\limits_{k,j,i=1}^{3+n_s} 
{\cal U}_{\beta k}W^\dagger _{kj}
e^{-i\frac{\tilde m_j^2}{2|\vec k|}L} 
W_{ji} {\cal U}^*_{\alpha i} .
\end{array}
\end{equation}
The $\lambda$ factors result from the normalization of the 
effective flavour states in Eq. (\ref{real}). 
If we ask for transitions of flavour neutrinos travelling 
long distances (allowing for heavy neutrinos to decay), 
these factors must 
be removed according to Eq. (\ref{lambda1})
\begin{equation}
A_{\nu _\alpha \to \nu _\beta} (L) = 
\lambda _\alpha \lambda _\beta
A_{\tilde \nu _\alpha \to \tilde \nu _\beta} (L) .
\end{equation}

For illustration we calculate the probability amplitudes for 
the case of 
2 standard families and  1 heavy neutrino.
We can as usual parametrise $\cal U$ and $\cal V$ in Eq. (\ref{Unu}) 
with 3 mixing angles and 1 phase  
\begin{equation}
{\cal U}= \left(
\begin{array}{cc}
c_{12} c_{13} & 
s_{12} c_{13} \\
&\\
-s_{12} c_{23}- c_{12} s_{23} s_{13}e^{i\delta}
& c_{12} c_{23}- s_{12} s_{23} s_{13}e^{i\delta} \\
& 
\end{array}
\right) \, , \label{uszur}
\end{equation}
\begin{equation}{\cal  V}=\left(
\begin{array}{c}
s_{13}e^{-i\delta} \\
s_{23} c_{13}
\end{array}
\right) \, ,
\label{eps}
\end{equation}
where $s_{ij}, c_{ij}$ stand for 
$\sin \theta_{ij}, \cos \theta_{ij}$, respectively, 
and $s_{13}$ and $s_{23}$ are small, with their products being 
constrained by Eq. (\ref{eps1}). 
($U_\nu$  has the same form as the mixing matrix for three 
families but now the third row corresponds to the mainly  heavy 
singlet neutrino, and the third column to the corresponding heavy 
mass eigenstate.) 
We can use the vacuum expressions to learn about the new effects.
Indeed, taking $W$ equal to the identity 
\begin{equation} 
P_{\nu_e \to \nu_e} (L)  =  | A_{\nu _e \to \nu _e} (L) |^2 = 
 c_{13}^4 \left(1-\sin ^2 2\theta _{12}\sin ^2\Delta\right), 
\end{equation}
with $\Delta = \frac{\Delta m_{21}^2 L}{4|\vec k|}$, and 
\begin{equation} 
\begin{array}{c}
P_{\nu_e \to \nu_\mu} (L)  =  c_{13}^2s_{13}^2s_{23}^2 
+  \sin 2\theta _{12} c_{13}^2 \left\{
s_{13}s_{23}c_{23}\sin \delta \sin 2\Delta \right. \\ 
\quad  +  \left. \left[ \sin 2\theta _{12} 
\left( c^2_{23}-s^2_{13}s^2_{23} \right) +
\cos 2 \theta _{12} s_{13} \sin 2 \theta _{23} \cos \delta
\right] \sin ^2\Delta \right\} .
\end{array}
\end{equation}
The sum of both probabilities is always smaller than 1. In 
fact if we also add the probability amplitude for producing the 
mainly heavy flavour eigenstate $P_{\nu_e \to \nu_N} (L)$ 
(which one may eventually detect through its decay products 
\cite{Aguila88}), we 
obtain $c_{13}^2$, which is smaller than 1 if the electron neutrino 
mixes with the heavy mass eigenstate, $s_{13} \neq 0$.
Besides, there are CP violating effects even with two families 
(or with three families and a vanishing mixing between the first 
and third one, or two degenerate light masses) 
\begin{equation}
\begin{array}{rcl} 
\Delta P^{CP}_{\nu_e \to \nu_\mu} (L) & =  
& P_{\nu_e \to \nu_\mu} (L) -  P_{\bar \nu_e \to \bar \nu_\mu} (L)  \\
& = &  c_{13}^2 \sin 2 \theta _{12} s_{13} \sin 2 \theta _{23}
\sin \delta \sin 2\Delta .
\end{array}
\end{equation}
At any rate, all new effects are suppressed by at least the product of 
two small mixings $s_{13}$ and/or $s_{23}$, and thus they are bounded by 
the stringent limits in Eq. (\ref{eps1}). Obviously we call the initial 
neutrino $e$ and the final $\mu$ but they stand for any two flavours. 
In fact the larger effects are expected for $\nu _\mu \to \nu _\tau$ 
transitions.

\section{Resonant oscillation of light neutrinos 
without heavy neutrino decoupling}
The same is true for neutrino oscillations in matter. 
For example in this case the usual resonant behaviour 
\begin{equation}
\sin 2\theta _{\mathrm {eff}} = 
\frac{\sin 2\theta_{12}}{\sqrt{(\frac{2\sqrt 2 G_F |\vec k|N_2}
{\Delta m^2_{21}}-\cos 2 \theta_{12})^2+ \sin ^2 2\theta_{12}}}
\end{equation}
writes
\begin{equation}
\sin 2\theta _{\mathrm {eff}} = c_{13}^2
\frac{A}{\sqrt{(B-\cos 2 \theta_{12})^2+ A^2 }},
\end{equation}
with 
\begin{equation}
\begin{array}{cl}
A^2 = &[\sin 2\theta _{12} + 
\frac{\sqrt 2 G_F |\vec k|N_n}{\Delta m_{21}^2}
s_{13}\sin 2 \theta _{23}\cos \delta]^2 
 \\ 
&+ (\frac{\sqrt 2 G_F |\vec k|N_n}{\Delta m_{21}^2})^2 
s^2_{13} \sin ^2 2 \theta _{23} \sin ^2\delta \  ,\\
B = &\frac{\sqrt 2 G_F |\vec k|(2N_e - N_n)}{\Delta m_{21}^2}c_{13}^2
+ \frac{\sqrt 2 G_F |\vec k|N_n}{\Delta m_{21}^2}
(c_{23}^2-s_{13}^2s_{23}^2) \ . 
\end{array}
\end{equation}
Thus the form is the same, but the resonance effect corresponds 
to values of the parameters corrected by amounts again suppressed 
by at least the product of 
two small mixings $s_{13}$ and/or $s_{23}$.
The important point is that the maximum $\sin 2\theta _{\mathrm {eff}}$ 
is not 1
but $c_{13}^2$ what gives another (difficult) way to measure the mixing 
between light and heavy neutrinos. 

\section{Conclusions}
Light neutrino masses are so small than mixing between light and heavy 
neutrinos must have a different origin if it is to be observable. 
This requires either fine tuning or models with two different heavy 
scales. Natural SM extensions realizing this scenario are 
$E_6$ models with two heavy scales of gauge symmetry breaking.
Models with extra dimensions have also typically two such scales, 
the compactification and the string scale. 

Independent of its origin one may wonder about the phenomenological 
implications of having heavy neutrinos with relatively large mixing 
with the SM ones. 
This case does not exhibit the cancellations present in the SM with 
only three massive light neutrinos but the departure from the SM 
predictions is bounded to be small, in fact smaller than the limits 
quoted in Eq. (\ref{eps1}). These bounds result from  
charged lepton processes highly suppressed in the SM. 
New heavy neutrinos manifest in these transitions through their 
interchange in loops; whereas in neutrino processes they show 
up at tree level. In any case it can be proven that the corrections 
involve at least two powers of the small mixing between light 
and heavy neutrinos. No such new effects have been observed, 
the required precision for their detection demanding improved 
measurements of rare charged lepton processes or neutrino 
experiments at a $\nu$ factory. In this case the main signature 
is the observation of CP violation together with no mixing 
between the first and third families. Other effects which are 
corrections to SM processes like the sum of probabilities 
not adding to 1 or modified resonance effects will be difficult to 
discriminate. At any rate the best place to look for is in 
$\mu$ and $\tau$ processes not involving $e$ 
because present limits are less stringent. 
Besides their masses are larger and it is generally believed  
that mixing effects have some kind of scaling with them, favouring 
the observation of SM departures in heavy flavours.

\section{Acknowledgments}
We thank J. Santiago for useful comments.
The work  was supported in part by the Polish Committee for 
Scientific Research under Grant 2P03B05418, 
MCYT under 
contract FPA2000-1558, Junta de Andaluc{\'\i}a group FQM 101 
and the European Community's Human Potential Programme under 
contract HPRN-CT-2000-00149 Physics at Colliders.

\end{document}